\documentclass[12pt]{article}
\usepackage[english]{babel}
\usepackage{a4wide}
\usepackage[ansinew]{inputenc}
\usepackage{amsfonts}
\usepackage{amsmath}
\usepackage{amssymb}
\usepackage{amsthm}
\usepackage{amstext}
\usepackage{newlfont}
\usepackage{graphicx}
\usepackage{color}

\title{M/g/c/c state dependent queueing  model for road traffic simulation}

\author{Nacira Guerrouahane$^{1}$\footnote{Corresponding author, e-mail: naciraro@hotmail.fr}~,
         Djamil Aissani$^{1}$, Louiza Bouallouche-Medjkoune$^{1}$\\ and  Nadir Farhi$^{2}$. \\~~ \\
      \small $^{1}$~LaMOS Research Unit, Faculty of Exact Sciences, University of Bejaia,\\ \small 06000 Bejaia,  Algeria.\\
      \small $^{2}$~Universit\'e Paris-Est, Ifsttar/Cosys/Grettia, F-77447,\\ \small Marne-la-Vall\'ee, Cedex 2, France. \normalsize }

\date{}

\begin{document}
\maketitle

\begin{abstract}
In this paper, we present a stochastic queuing model for the road traffic, which captures the stationary density-flow
relationships in both uncongested and congestion conditions.
The proposed model is based on the $M/g/c/c$ state dependent queuing model of Jain and Smith, and
is inspired from the deterministic Godunov scheme for the road traffic simulation.
We first propose a reformulation of the $M/g/c/c$ state dependent model that works with density-flow fundamental diagrams
rather than density-speed relationships. We then extend this model in order to consider upstream traffic demand as well as
downstream traffic supply. Finally, we calculate the speed and travel time distributions for the $M/g/c/c$ state dependent
queuing model and for the proposed model, and derive stationary performance measures (expected number of cars, blocking
probability, expected travel time, and throughput). A comparison with results predicted by the $M/g/c/c$ state dependent
queuing model shows that the proposed model correctly represents the dynamics of traffic and gives good performances
measures. The results illustrate the good accuracy of the proposed model.

\end{abstract}

\section{Introduction}
\label{intro}
Traffic flow on freeways is a complex process with many interacting components and random perturbations such
as traffic jams, stop-and-go waves, hysteresis phenomena, etc. These perturbations propagate from upstream to downstream sections.
During traffic jams, drivers are slowing down when they observe traffic congestion in the downstream section, causing upstream
propagation of a traffic density perturbation.

Models for flow on a link have developed from the fundamental diagram, where flow is a function of density via the macroscopic LWR first-order continuum model
( i.e., Lighthill-Whitham-Richards theory of kinematic waves)~\cite{Lig55,Ric56}.
In this paper, we propose a stochastic traffic model based on the queuing model of~\cite{Smi94} and on the Godunov scheme~\cite{God59,Leb96} of the LWR
traffic model. We calculate a stationary probability distribution of the $M/g/c/c$ state dependent queuing model~\cite{Smi97,Smi94} on a road section, by
considering density-flow fundamental diagrams rather than density-speed ones~\cite{ICNAAM15}. The model suppose a triangular fundamental diagram which correctly captures
the stationary density-flow relationships in both uncongested and congestion conditions. In this, we use the functions of traffic demand and supply for
the section, and derive a model for a road with a downstream supply,  we present stationary performance measures (expected travel time, throughput, etc.), and we derive a distributions of speed and travel time. The model we propose here can also be used to the analysis of travel times through road traffic~\cite{Far08,Far14a,Far14b}.

The remainder of this paper is organized as follows. In Section 2, we first present a review of the existing works on
literature. In this regards, we present a short review of the $M/g/c/c$ state dependent queuing model of Jain and Smith. In Section 3, we
rewrite the $M/g/c/c$ state dependent queuing model on a road section by considering density-flow fundamental diagrams rather than density-speed ones
(triangular fundamental diagram). In Section 4, we consider the traffic demand and supply functions for the section, and derive a model for a road with
a downstream supply. We derive some performance measures (expected number of cars, blocking probability, expected travel time and throughput) and we
compared it by $M/g/c/c$ state dependent queuing model . In Section 5 , we calculate the speed and expected travel time distributions, and we present
our simulation results. In Section 6, we briefly summarize our findings.

  \section{Literature review} 
  \label{sec1}

The dynamics of traffic flows in road networks is complex, and is subject to stochastic disturbances. Congested networks involve complex traffic
interactions. Providing an analytical description of these intricate interactions is challenging.
The study of network congestion is of interest in various fields, ranging from the analysis
of spillbacks (i.e. the backwards propagation of congestion) in urban traffic or pedestrian
traffic~\cite{Smi97,Smi94}.

In this article, we are concerned with two traffic models: the M/G/C/C state-depentent queueing static model and the LWR (Lighthill-Whitham-Richards) 
first order dynamic model~\cite{Lig55,Ric56}. Numerical schemes for the LWR model have been performed since decades \cite{God59}, \cite{Dag94}, \cite{Leb96}.
There has been a recent interest in the development of stochastic link models. Most studies have
considered stochastic cell-transmission models~(CTM)~\cite{Dag94}, where traffic demand and supply functions are used.
In~\cite{Oso11}, the authors proposed a stochastic formulation of the link-transmission model, which is an operational instance of Newell’s theory of kinematic waves~\cite{Newel93}. The kinematic wave model~(KWM) is more recently developed~\cite{Tam10}.
In~\cite{Boel06}, the compositional stochastic model extends the cell transmission model~\cite{Dag94} by defining sending (demand) and receiving (supply) functions explicitly as random variables.

Several simulation models based on queueing theory have been developed, but few studies have explored the potential of the queueing theory framework to
develop analytical traffic models. In~\cite{Oso11}, the authors proposed an analytical stationary model, which is directly derived from the KWM.
A review of stationary queueing models for highway traffic and exact analytical stationary queueing models of signalized and unsignalized intersections are given by several authors~\cite{HeiO1,Van08,Vandael00}. In~\cite{Guerr14,Hei97}, the authors contributed to the study of signalized intersections and presented a unifying approach to both signalized and unsignalized intersections. These approaches resort to infinite capacity queues, and thus fail to account for the
occurrence of breakdown and their effects on upstream links.

Calculus of traffic flow breakdown probability remains an important issue when analyzing the stability and reliability of transportation
system ~\cite{Birlon05,Break10}.
Finite capacity queueing network model~(FCQN) are of interest for a variety of applications such as the study of
manufacturing networks, circulation systems and prison networks~\cite{Smi94}, etc. FCQN model allows to account for finite lengths, which enables the modeling and analysis of breakdowns.
The methods proposed in~\cite{Bed12,Smi97,Smi94} resort to finite capacity queueing theory and derive stationary performance measures.
In~\cite{cruz07}, the authors describe a methodology for approximate analysis of open state dependent $M/g/c/c$ queueing networks.
The evacuation problem was analyzed using $M/g/c/c$ state dependent queuing networks \cite{cruz05,Yuhaski} when an algorithm was proposed to optimize the
stairwell case and increase evacuation times towards the upper stories. In~\cite{ICNAAM15}, the authors proposed a reformulation of the linear case of $M/g/c/c$ state queueing model~\cite{Smi97}, which uses the density-flow fundamental diagrams and consider upstream traffic demand and downstream traffic supply.

\section{Review on M/g/c/c state dependent queuing model}
\label{sec2}

In this section, we present a  review on the $M/g/c/c$ state dependent queuing model of Jain and Smith. This
model was used to model pedestrian and vehicular traffic flow~\cite{Smi97,Smi94}.

A link of a road network is modeled with $c$ servers set in parallel, where $c$ is the link capacity (the maximum number of occupants in the link).
The model assumed that the average speed $v_n$ depends on the number of occupants $n$ on the road, according to a non-increasing density-speed relationship.

In accordance to Tregenza's empirical studies~\cite{Tregenza}, the average speed that an occupant will move through a link depends on several factors but mainly is a function of the number of occupants in the link. Based on these studies, linear and exponential congestion models are developed for the average pedestrian/vehicles speed in traffic links~~\cite{Smi97,Yuhaski}.
The linear congestion model is based on the idea that the service rate is a linear function of the number of occupants
in the link and is given as follows.

\begin{equation}\label{eq-vn}
   v_n = v_{f} \frac{(c-n+1)}{c},
\end{equation}

The exponential congestion model is based on the idea that the service rate is related to the number of occupants by an exponential function and is given as follows.

\begin{equation}\label{eq-ve}
  v_n = v_{f} \exp\left[-\left(\frac{n-1}{\beta}\right)^{\gamma}\right].
\end{equation}

  \begin{figure}[htbp] \label{fig-approximation}
  \begin{center}
      \includegraphics[width=0.5\textwidth]{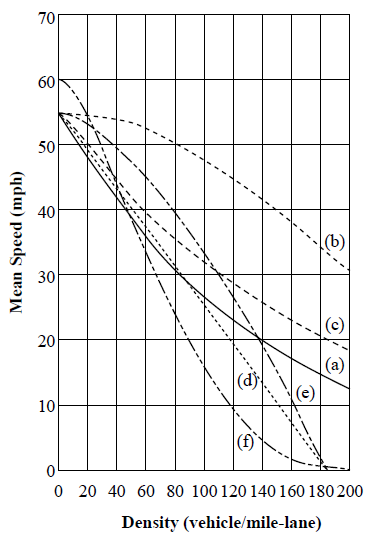}%
     \caption{Exponential approximation of empirical M/g/c/c state-dependent traffic flow model, (a) (Jain and Smith, 1997), and
      empirical distributions for vehicular traffic flows, (b) (Edie, 1961), (c) (Underwood, 1961), (d) (Greenshields, 1935), (e) (Drake et al., 1967), and (f) (Transportation Research Board, 2000)~\cite{Smi97}.}
      \end{center}
\end{figure}

In~(\ref{eq-ve}), $\beta$ and $\gamma$ are shape and scale parameters respectively. Parameters $\beta$ and $\gamma$ are
found by fitting points to the
curve in Fig. 1.  In fact,  Fig.1 presents an approximation of empirical vehicular speed-density curves, based on various empirical studies~\cite{Smi97}. Fitting the points $(1, v_{f})$, $(a, v_{a})$ and $(b, v_{b})$ gives one the algebraic relationships shown below~\cite{Yuhaski}.
$$ \begin{array}{ll}
       \beta & = \frac{a-1}{\left[\ln(v_{f}/v_{a})\right]^{1/\gamma}}=\frac{b-1}{\left[\ln(v_{f}/v_{b})\right]^{1/\gamma}}, \\
             & \\
       \gamma & = \ln \left[\frac{\ln(v_{a}/v_{f})}{\ln(v_{b}/v_{f})}\right]/\ln\left(\frac{a-1}{b-1}\right).
   \end{array}$$
The values $a$ and $b$ are arbitrary points used to fit the exponential curve.
In vehicular related applications~\cite{Smi97}, commonly used values are $a =~20Lk$ and $b =~140Lk$ corresponding to densities of 20 and
140 veh/mi-lane respectively. Looking at the curves presented in Fig. 1, reasonable values for such points are $v_{a} =~ 48$ miles per hour
and $v_{b} =~ 20$ miles per hour. $L$ is the  length of the link and  $k$ is its width (or number of lanes).

The arrival process of cars into the link is assumed to be Poisson with rate $\lambda$,
while the service rate is general and depend on the number of occupants $n$ on
the link. A normalized service rate $f(n)$ is defined as the ratio of average speed to free speed, in order to capture
congestion effects, and is taken $f(n) = v_{n}/v_{f}$, $0 \leq f(n) \leq 1$. In the linear case, we have
$f(n) = (c-n+1)/c$. In the exponential case, we have $f(n) = \exp[-((n-1)/\beta)^{\gamma}]$.

The stationary probability distribution $P_n = P(N = n)$ of the number of occupants $N$ in the $M/g/c/c$ state dependent model have been
developed in~\cite{Yuhaski} and shown in~\cite{Smi94} to be stochastic equivalent to a pure Markovian $M/M/c/c$ queueing model. Then, these probabilities can be written as follows.
\begin{equation}\label{eq1}
\begin{array}{ll}
 P_n = \frac{(\lambda L/v_{f})^{n}}{\prod_{i=1}^{n}i f(i)} P_0, \qquad n=1,..,c.\\ \\
 P_0 = \left(1+\sum_{n=1}^{c}\frac{(\lambda L/v_{f})^{n}}{\prod_{i=1}^{n}i f(i)}\right)^{-1}.
 \end{array}
\end{equation}
where $L$ is the length of the link section and $v_{f}$ is the speed corresponding to one occupant in the link (ie. the free speed).

From $P_n$,  we can easily derive important performance measures.
\begin{itemize}
    \item The blocking probability :\\$ P_{c} = P_{0} (\lambda L/v_{f})^c / \prod_{i=1}^{c} i f(i),$
     \item The throughput : $ \theta = \lambda(1-P_{c}),$
     \item The expected number of cars in the section : $ \bar{N} = \sum_{n=1}^{c}n P_{n},$
     \item The expected service time : $W = \bar{N} / \theta.$
  \end{itemize}

\section{Model of road section}
\label{sec3}

In this section, we slightly modify the $M/g/c/c$ state dependent queueing model of Jain and Smith, by defining the normalized service rate $f(n)$ as the ratio of the average flow ($q_n$) by the maximum flow ($q_{\max}$), rather than the average speed ($v_n$) by the free speed ($v_f$).
This modification will permit us to consider the demand and supply functions of a road section, and then to use them in the case
where two or many sections are set in tandem.
In the following, we present the $M/g/c/c$ state dependent queuing model on one road section, for which we consider a triangular fundamental
traffic diagram~\cite{ICNAAM15}.
\begin{equation}\label{tfd}
   Q(\rho) = \min(v_{f} \rho, w(\rho_{j}-\rho)).
\end{equation}

The demand and the supply functions $\Delta(\rho)$ and $\Sigma(\rho)$ respectively are given as follows.
\begin{align}
   & \Delta(\rho) = \min(v_f \rho, q_{\max}), \label{dem-fun} \\
   & \Sigma(\rho) = \min(q_{\max}, w(\rho_j - \rho)). \label{sup-fun}
\end{align}

where $q_{\max} = \rho_j / (1 / v_f + 1 / w)$, and $\rho = n / L$.\\
$\rho, Q(\rho), v_f, w$, $\rho_j$, $L$, $q_{\max}$ and $n$ denote respectively the car-density in the road section, the car-flow, the free speed,
the backward wave speed, the jam-density, the length of the road section, the maximum flow and the number of cars.

We define $T$ as the service time, ie the time needed for a car to pass through one road section. Let us notice here that the service depends on both traffic demand and traffic supply, since we have here a state-dependent service model. The expected service time $E(T)$ depends on the number
$n$ of cars on the road and is given by $E(T)=L / v_{n}$, where $v_{n}= Q(\rho) / \rho.$

The expected service time $E(T)$ is then given as follows.
\begin{equation*}
E(T)=\frac{n}{\min(v_{f}\frac{n}{L}, w(\frac{c-n}{L}))}.
\end{equation*}

The average service rate $\mu$ of one server (one car place) is then given as follows.
\begin{equation*}
\mu=\frac{1}{E(T)} = \frac{v_{n}}{L} = \frac{\min(v_{f}\frac{n}{L}, w(\frac{c-n}{L}))}{n}.
\end{equation*}

The overall service rate $q_n$ of the road section (the queueing system) with $n$ cars is
equivalent to the number of occupied servers multiplied by the rate of each server, and is nothing but the car-flow on the section,
given by the fundamental diagram~(\ref{tfd}).
\begin{equation*}
 q_{n}= n\mu = \min\left(v_{f}\frac{n}{L}, w\left(\frac{c-n}{L}\right)\right).
\end{equation*}
The normalized service rate is then fixed to
\begin{equation}\label{fn}
   f(n) = \frac{q_{n}}{q_{\max}} = \frac{\min(v_{f} \frac{n}{L}, w(\frac{c-n}{L}))}{q_{\max}}.
\end{equation}
We have $0 \leq f(n) \leq 1$.

Stationary probabilities of the number of cars on the road section are derived by
substituting the expression of $q_n$, into the Chapman-Kolmogorov equations for solving the probabilities
of a single queue \cite{Smi94}.
\begin{equation}\label{eq}
\begin{array}{ll}
    P_{n} = \frac{\lambda^{n}}{\prod_{i=1}^{n}\mu_{i}} P_{0},\qquad n=1,..,c.\\
    P_{0} = \left(1+\sum_{n=1}^{c}\frac{\lambda^{n}}{\prod_{i=1}^{n}\mu_{i}}\right)^{-1}.
\end{array}
\end{equation}

Then, the stationary probability distribution of the number of cars on the road section is given as follows~\cite{ICNAAM15}
\begin{equation}\label{eq2}
\begin{array}{ll}
     P_n = \frac{(\lambda)^{n} (\frac{L}{v_{f}})^{n_{cr}} (\frac{L}{w})^{n-n_{cr}}}
          {\prod_{i=1}^{n_{cr}}i\prod_{i=n_{cr}+1}^{n} (c-i)} P_{0}, \qquad n=1,..,c.\\ \\
     P_0 = \left(1+\sum_{n=1}^{c}\frac{(\lambda )^{n} (\frac{L}{v_{f}})^{n_{cr}} (\frac{L}{w})^{n-n_{cr}}}
          {\prod_{i=1}^{n_{cr}}i\prod_{i=n_{cr}+1}^{n} (c-i)}\right)^{-1}.
\end{array}
\end{equation}
where $n_{cr}=\rho_{cr}L$ is the number of cars corresponding to the critical car-density.

\subsection{Model with a downstream supply}
\label{Model downstream}

We model here a road section with $M/g/c/c$ state dependent queuing model, as presented above, but we consider that the service of section 1 is
constrained by the supply flow of the downstream section (section 2), as in Fig. 2.

\begin{figure}[htbp]
\begin{center}
    \includegraphics[width=0.7\textwidth]{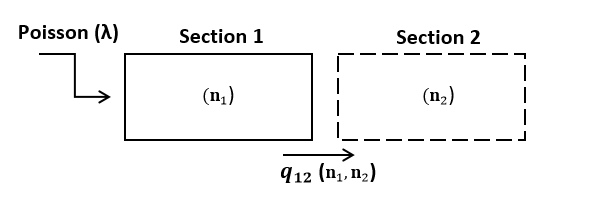}
    \caption{A road section with a downstream supply constraint.}
    \end{center}
    \label{tandem}
\end{figure}

The model assume that the supply flow of the downstream section is stochastic, the stationary probability distribution of the number of cars in the downstream section is given and a triangular fundamental traffic diagrams for the two sections are given ( See Fig. 3).
\begin{figure}[htbp]{\label{diagramme}}
\begin{center}
    \includegraphics[width=0.86\textwidth]{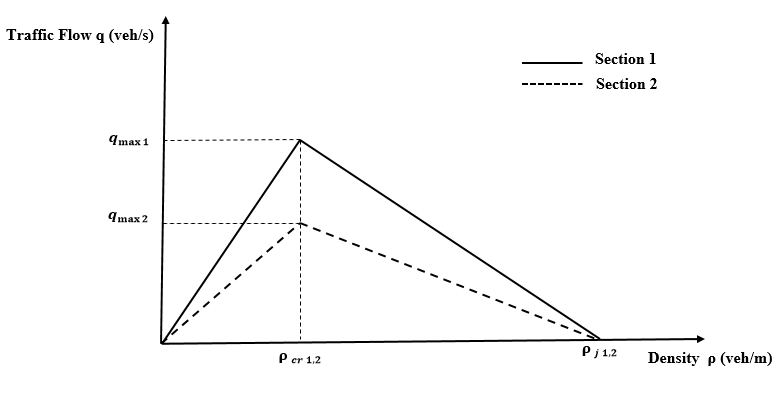}
    \caption{Triangular fundamental diagrams.}
\end{center}
\end{figure}

The car-flow outgoing from section 1 and entering to section 2 is assumed to be given by the minimum between the traffic demand on section 1 and the traffic supply of section 2.
\begin{equation*}
q_{12}(n_1,n_2) = \min(\Delta_1(\rho_1), \Sigma_2(\rho_2))
                = \min\left(v_{f1}\frac{n_1}{L_1}, q^{\max}_1, q^{\max}_2, w_2\left(\frac{c_2-n_2}{L_2}\right)\right).
\end{equation*}
Therefore, the normalized service rate $f(n_1,n_2)$ of section 1 is given as follows.
\begin{equation}
f(n_1,n_2) = \frac{q_{12}}{q^{\max}_1}=  \frac{\min\left(v_{f1} \frac{n_1}{L_1}, q^{\max}_1, q^{\max}_2, w_2 (\frac{c_2-n_2}{L_2}) \right)}
               {q^{\max}_1}.
\end{equation}

\begin{enumerate}
\item The stationary probability distribution of the number of cars on section 2, is assumed fixed and given by~(\ref{eq2}),
in function of $\theta$, as follows.

\begin{equation}\label{eq5}
  \begin{array}{ll}
    P_{n_2}^{(2)} (\theta) &  = \frac{(\theta)^{n_2} (\frac{L_{2}}{v_{f2}})^{n_{cr_2}} (\frac{L_2}{w_2})^{n_{2}-n_{cr_2}}}
          {\prod_{i=1}^{n_{cr_2}}i\prod_{i=n_{cr_1}+1}^{n_2} (c_{2}-i)} P_{0} (\theta) ,\qquad n_2=1,..,c_2. \\ \\
         P_0^{(2)} (\theta) & = \left(1+\sum_{n_{2}=1}^{c_2}\frac{(\theta )^{n_2} (\frac{L_2}{v_{f2}})^{n_{cr_2}} (\frac{L_2}{w_2})^{n_{2}-n_{cr_2}}}
          {\prod_{i=1}^{n_{cr_2}}i\prod_{i=n_{cr_2}+1}^{n_{2}} (c_{2}-i)}\right)^{-1}.
  \end{array}
\end{equation}

\item In section 1, we have an $M/g/c_1/c_1$ system, parameterized by the traffic supply of the downstream section (the number of cars in section 2).
\begin{itemize}
\item The stationary probability distribution of the number of cars on section 1,
parameterized by the number of cars on section 2, is given as follows.

\begin{equation}\label{eq3}
  \begin{array}{l}
     P_{n_1 \mid n_2} (\lambda) = \frac{(\lambda / q^{\max}_1)^{n_1}}{\prod_{i=1}^{n_1} f(i, n_2)} P_{0\mid n_2}(\lambda), \\
     P_{0\mid n_2} (\lambda) = \left(1+\sum_{n_1 = 1}^{c_1} \frac{(\lambda / q^{\max}_1)^{n_1}}{\prod_{i=1}^{n_1} f(i, n_2) }\right)^{-1}.
  \end{array}
\end{equation}

\item Then, the stationary probability distribution of the number of cars on section 1 is obtained as follows.
\begin{equation}\label{eq4}
 \begin{array}{l}
    P^{(1)}_{n_1} (\lambda, \theta) = \sum_{n_2 = 0}^{c_2} P_{n_1\mid  n_2} (\lambda) P^{(2)}_{n_2} (\theta), \qquad n_1=1,..,c_1. \\\\
    P^{(1)}_{0} (\lambda, \theta) = \sum_{n_2 = 0}^{c_2} P_{0\mid  n_2} (\lambda) P^{(2)}_{n_2} (\theta).
  \end{array}
\end{equation}
\end{itemize}
\item The average outflow from section 1, $\theta$, is given as follows (using little's law).
      \begin{equation}\label{theta}
	\theta= \lambda \left( 1 - P^{(1)}_{c_1}(\lambda, \theta)\right).
      \end{equation}
\end{enumerate}

Using the data of Table~1, Fig. 4 compares the stationary probability distribution of the number of cars $n$ on the road section for our model with downstream supply (formula~\ref{eq4}), with red color and for the linear case of Jain and Smith model (formula~\ref{eq1}), with blue color. The arrival rates $\lambda$ is varied from one illustration to another ($\lambda = 0.5$~veh/s,~$1$~veh/s and $~1.5$~veh/s).

\begin{figure}[htbp] \label{fig-distribution}
  \includegraphics[width=0.32\textwidth]{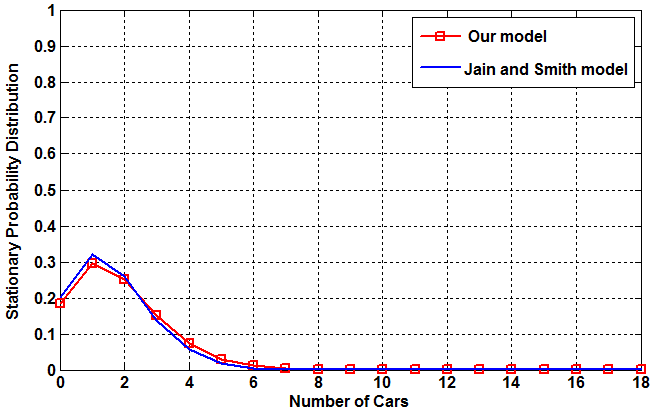}
  \includegraphics[width=0.32\textwidth]{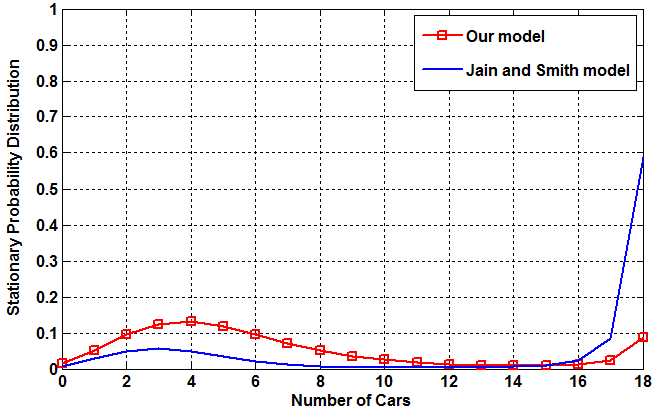}
  \includegraphics[width=0.32\textwidth]{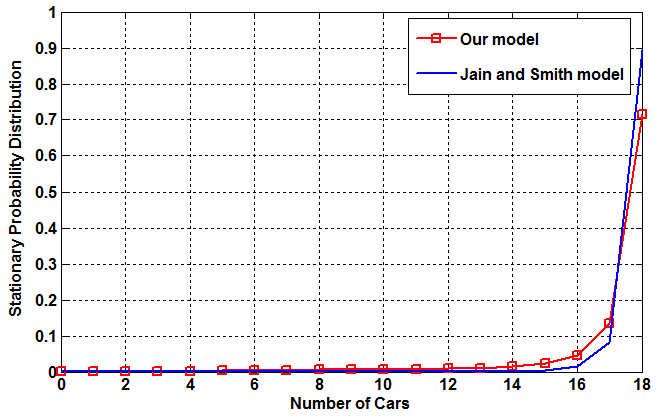}\\
  \caption{Stationary probability distributions. $\lambda =$ $0.5$ veh/s, $1$ veh/s and $1.5$ veh/s respectively from left to right sides.}
\end{figure}

\begin{figure}[htbp] \label{fig-blocage}
\includegraphics[width=0.5\textwidth]{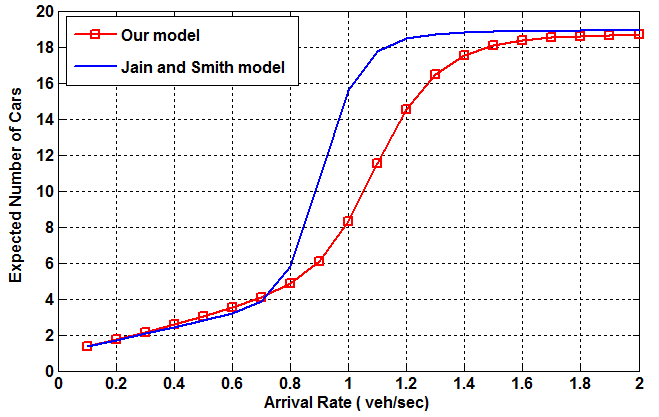}
  \includegraphics[width=0.5\textwidth]{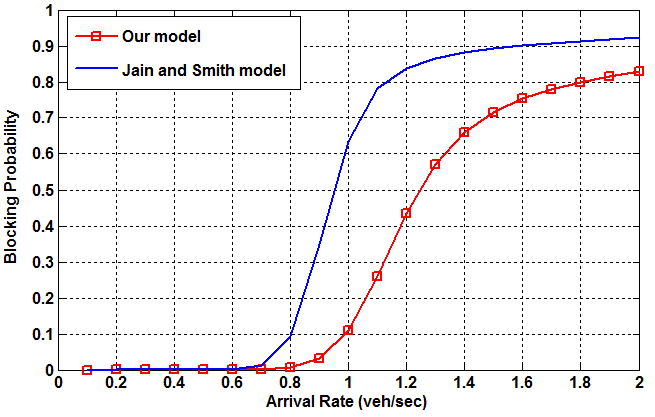}
    \caption{Comparison of the expected number of cars, in left. Comparison of the blocking probability, in right.
     Our model with red color. Jain and Smith model with blue color.}
\end{figure}

Fig. 4 shows that the number of cars $n$ on the road section increases, in term of stationary probability, with the arrival rate $\lambda$.
The difference between the two stationary distributions in the middle of Fig. 4 ($ \lambda = 1$~veh/s) can be explained by using the flow ratio rather than the speed ratio as a measure of the service rate. Moreover, in the case of Jain and Smith model (blue color), the probability of saturation of the road section is larger than our model. Fig. 5 shows for an increasing arrival rate $\lambda$, the expected number of cars in the road section ($\bar{N} = \sum_{n=1}^{c} n P_n$) and the blocking probability ($P_{c}= P(N = c)$), for our model with downstream supply (red color) and for the model of Jain and Smith (blue color).

  \subsection{Performance measures}

In the following, we give an illustration example. We consider two road sections as in Fig. 2. We assume that the fundamental diagrams
for those roads are triangular, see Fig. 3. Table~1 gives the parameters for the illustrations.

\begin{table}[htbp]
\begin{center}
  \begin{tabular}{c||c|c|c|c|c|c|c}
  \hline
     Section $i$ & $L$ (m) & $v_{f}$ (m/s) & $w$ (m/s) & $\rho_j$ (veh/m) & $q_{\max}$ (veh/s) & $\rho_{cr}$ (veh/m) & $c$ (veh) \\
     \hline
     1 & 100 & 28 & 14 & 0.18 & $1.6$ & 0.06 & 18 \\
     \hline
     2 & 100 & 14 & 7 & 0.18 & $0.8$ & 0.06 & 18\\
     \hline
  \end{tabular}
  \begin{center}
  \caption{Sections parameters.}
  \end{center}
\end{center}
\end{table}

\begin{itemize}
 \item[-] \textbf{\emph{Expected travel time (W)}}
 \end{itemize}

The travel time through a road section is a random variable and is a function of the number of cars on the road section.
Since the road section has a finite length, it can be seen as a queuing system with a finite capacity, for which
the travel time $W$ (or, service time) can be derived using the Little's law. The latter law gives the travel time as the average number of cars in the road ($\bar{N}$) divided by the effective arrival rate ($\lambda (1-P_c)$).
\begin{figure}[htbp] \label{fig-service}
\begin{center}
\includegraphics[width=0.5\textwidth]{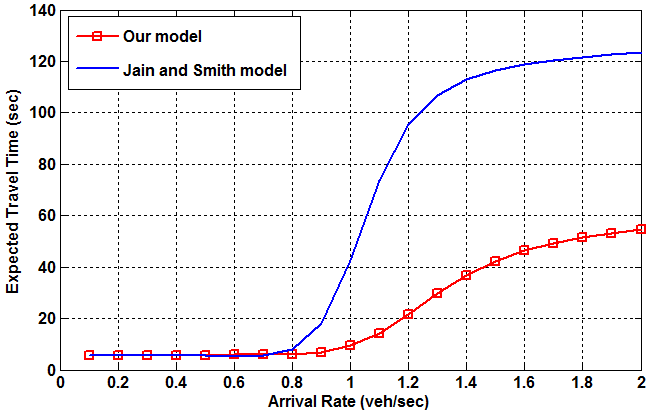}
   \caption{Comparison of the expected travel time through section 1. In red color, our model. In blue color, model of Jain and Smith.}
\end{center}
\end{figure}

Fig. 6 compares, for an increasing arrival rate $\lambda$, the expected travel time through road section 1 of Fig. 2, for our
model with downstream supply and for the linear case of Jain and Smith model (see Table 1 for the parameters
of the road section).

The curves of the expected travel time for the two models are not monotone increasing. There is an upper limit for the arrival rate $\lambda$, which is $0.8$~veh/s, from there the vehicles begins to slow down and the expected travel time begins to increase.
Therefore, the expected travel time before this point ($\lambda =$ 0.8~veh/s) is very low (around the free time ($L/v_{f}$)) for the linear case of Jain and Smith model and our model. When the traffic volume is large ($\lambda$ $>$ 0.8~veh/s) vehicles will slow down and the expected travel
time increase in value for the two models of traffic, but still lower in our model with downstream supply. The expected travel time continues to increase with demand beyond capacity.


\begin{itemize}
 \item[-] \textbf{\emph{Throughput ($\theta$)}}
 \end{itemize}

Throughput $\theta$ can be calculated using two methods. Measuring the effective arrival rate of the accepted cars in the system
$(\theta = \lambda(1-P_{c}))$. Measuring the effective departure rate of the served cars in the system
$(\theta = \sum_{n=1}^{c} q_n P_{n})$.
\begin{figure}[htbp]
\begin{center}
\includegraphics[width=0.5\textwidth]{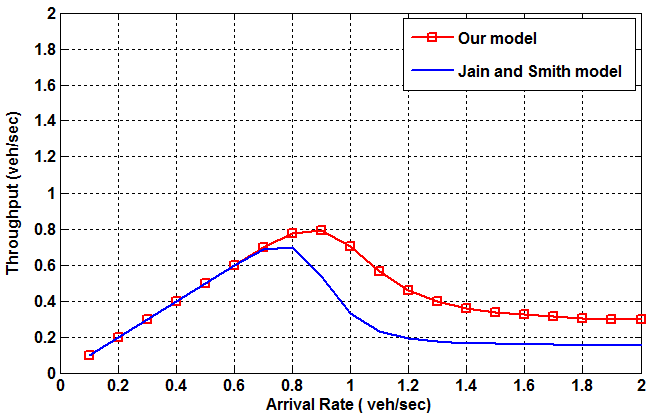}
   \caption{Comparison of the throughput through section 1. In red color, our model. In blue color, model of Jain and Smith.}
\end{center}
\end{figure}

Figure~7 compares, for an increasing arrival rate $\lambda$, the throughput through road section 1 of Fig. 2, for our model with downstream supply and for the linear case of Jain and Smith model (see Table 1 for the parameters of the road section).

The curves of throughput increases linearly with arrival rate $\lambda$, but there is a halt to the monotonic increase. The throughput decreases with the arrival rate from the value $\lambda = 0.8$ veh/s, which corresponds to the flow capacity of the second section (minimum between flow capacities of sections 1 and 2).

When the blocking probability is low,  the throughput is linear up to $0.8$ veh/s. From that arrival rate ($\lambda = 0.8$ veh/s), the blocking probability increases (see Fig.5 in right), the throughput decreases, the expected number of cars also increases up to the system capacity $c$ (see Fig.5 in left), and the expected travel time worsens around $10$ times.
Arrival rates above $0.8$ cannot improve the system throughput that reaches its limit around $\theta = 0.32$ veh/s for our model and $\theta = 0.17$ veh/s for Jain and Smith model. Then, the system would be able to give a higher throughput under an arrival rate of $0.8$ veh/s. That seems to be the flow capacity of the second section.

\section{Speed and travel time distributions}

One of the most basic formula in traffic flow theory is the one expressing the interdependence of the average car-flow $(q)$, the average car-density
$(\rho)$ and the average car-speed $(v)$. The formula tells that $q = \rho v$.
When two of the three variables are known, the third variable can easily be obtained.

The average car-speed $v_n$ through a road section is a random variable because the number of cars $n$ on the road section is random.
Using the expression of the linear speed in the model of Jain and Smith~(equation~(\ref{eq-vn})), the car-speed probability distribution is given by:
\begin{equation*}
P(v_{n}=v)=P\left(n=\left\lfloor 1+c\left(1-\frac{v}{v_{f}}\right)\right\rfloor\right).
\end{equation*}
in which $\lfloor x \rfloor$ is the largest integer not superior to $x$. Then, the car-speed distribution is given as follows.
\begin{equation}\label{eq7}
\begin{array}{ll}
    P_{v} & =  P(v_{n}=v) = \frac{(\lambda L/v_{f})^{\lfloor1+c(1-v/v_{f})\rfloor}}{\prod_{i=1}^{\lfloor1+c(1-v/v_{f})\rfloor}i (c-i+1)/c} P_{0}, \qquad v=1,..,v_{f}.\\ \\
    P_{0} & = \left(1+\sum_{v=1}^{v_{f}}\frac{(\lambda L/v_{f})^{\lfloor1+c(1-v/v_{f})\rfloor}}{\prod_{i=1}^{\lfloor1+c(1-v/v_{f})\rfloor}i (c-i+1)/c}\right)^{-1}.
     \end{array}
\end{equation}

The average travel time $\tau$ through a road section can be evaluated given the road section length $L$ and the average car-speed $v$.
Basically, we have $\tau = L/v$. By this, the travel time probability distribution is given by:
\begin{equation*}
 P(\tau=t) = P(v=\frac{L}{t}) = P\left(n=\left\lfloor1+c\left(1-\frac{L}{t v_{f}}\right)\right\rfloor\right).
 \end{equation*}

Then, the average travel time distribution is given as follows.
\begin{equation}\label{eq8}
\begin{array}{ll}
 P_{t} & =   P(\tau=t) = \frac{(\lambda L/v_{f})^{\lfloor1+c(1-L/(t v_{f}))\rfloor}}{\prod_{i=1}^{\lfloor 1+c(1-L/(t v_{f}))\rfloor} i (c-i+1)/c } P_{0},
 \qquad t= \lfloor L/v_{f} \rfloor,..,L. \\ \\
 P_{0} & = \left(1+\sum_{t= \lfloor L/v_{f} \rfloor}^{L}\frac{(\lambda L/v_{f})^{\lfloor1+c(1-L/(t v_{f}))\rfloor}}{\prod_{i=1}^{\lfloor1+c(1-L/(t v_{f}))\rfloor}i (c-i+1)/c}\right)^{-1}.
 \end{array}
\end{equation}

Using the parameters of section 1 in Table 1. Fig. 8 shows the histograms for the probability distribution of the average car-speed and the average travel time through the road section, for the linear case of the model of Jain and Smith. The arrival rate considered is $\lambda = 0.8$ veh/s.

\begin{figure}[htbp]\label{speed-travel-mgcc}
  \includegraphics[width=0.51\textwidth]{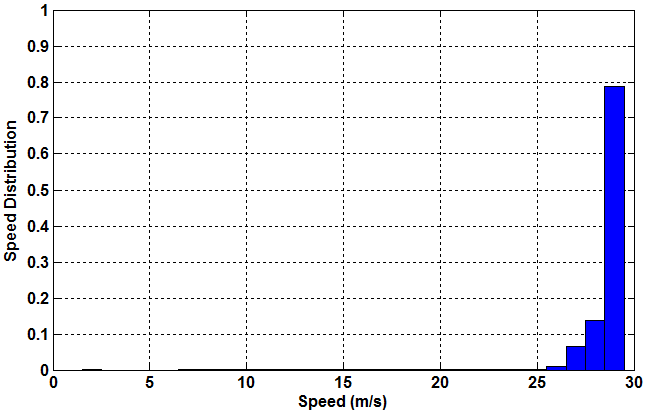}
  \includegraphics[width=0.53\textwidth]{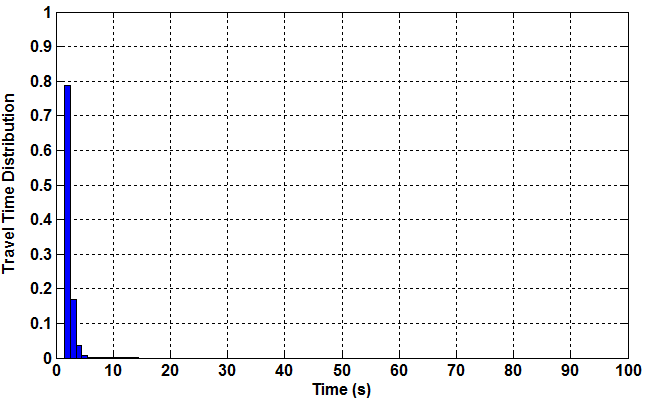}
  \caption{Car-speed probability distribution histogram, in left. Average travel time distribution histogram, in right. Linear case of Jain and Smith model.
     The parameters of the road section are those of road section 1 in Table 1, and the arrival rate is $\lambda = 0.8$ veh/s.}
\end{figure}

For our road section model with a triangular fundamental diagram, the average car-speed $v_n$ is given by the car-flow ($Q(\rho)$) in
the road  divided by the car-density ($\rho$).\\

The car-speed probability distribution satisfies.
\begin{equation*}
P(v_n = v) = P\left(\frac{\min(v_{f}\frac{n}{L}, w(\frac{c-n}{L}))}{\frac{n}{ L}} = v\right).
\end{equation*}

Then, two cases are distinguished:

\begin{enumerate}
  \item $\rho \leq \rho_{cr} \Longrightarrow v_n = v_f$. Then, $ P(v_n = v_f) = \sum_{n=0}^{n_{cr}} P(N=n),$
  \item $\rho > \rho_{cr}   \Longrightarrow  v_n < v_f$. Then, $P(v_n = v) =  P\left( N =\lfloor\frac{w c }{v + w}\rfloor\right).$
\end{enumerate}

The car-speed probability distribution is then given as follows.
\begin{equation}
 P(v_n = v) =
\begin{cases}
              0 & \text{if}~~ v > v_{f},\\
              \sum_{n=0}^{n_{cr}} P(N=n) & \text{if }~~ v = v_{f},\\
              P\left( N =\lfloor\frac{w c}{v + w}\rfloor \right) & \text{if}~~ v < v_{f}.

\end{cases}
\end{equation}

Similarly, we get the following formula for the probability distribution of the average travel time $\tau$ through a road section.
We use the formula $\tau = n/q$.

$$P(\tau = t) = P\left(\frac{n}{\min(v_{f} n / L, w((c-n)/L))} = t\right).$$

Then, two cases are distinguished:
\begin{enumerate}
  \item $\rho \leq \rho_{cr}\Longrightarrow \tau= L/v_f$. Then, $P(\tau = L/v_f) = \sum_{n=0}^{n_{cr}} P(N=n)$,
  \item $\rho > \rho_{cr} \Longrightarrow \tau > L/v_f$. Then, $P(\tau = t) = P\left( N=\lfloor \frac{w c t }{L + w t}\rfloor\right)$.
\end{enumerate}

The average travel time probability distribution is then given as follows.
\begin{equation}
P(\tau = t) =
  \begin{cases}
    0 & \text {if}~~ t<L/v_{f}, \\
    \sum_{n=0}^{n_{cr}} P(N=n) & \text{if} ~~t = L/v_{f}, \\
    P\left( N = \lfloor\frac{w c t}{L + w t}\rfloor\right) & \text{if}~~ t> L/v_{f}.
  \end{cases}
\end{equation}

Fig. 9 displays the histograms for the probability distribution of the average car-speed and the average travel time through the road section, for our
model with downstream supply. The arrival rate is fixed to $\lambda= 0.8$~veh/s. Fig. 9 shows that when the arrival rate is low, speed distribution corresponds to the free speed $(v_{f} = 28$~m/s ) with a high probability, and the average travel time distribution corresponds to the free time $(t_{f} = L/v_{f} = 3.5$ s), with the same probability. In this case, traffic is fluid because the section is not occupied (or blocked) by the cars.

\begin{figure}[htbp]\label{fig-compar2}
  \includegraphics[width=0.51\textwidth]{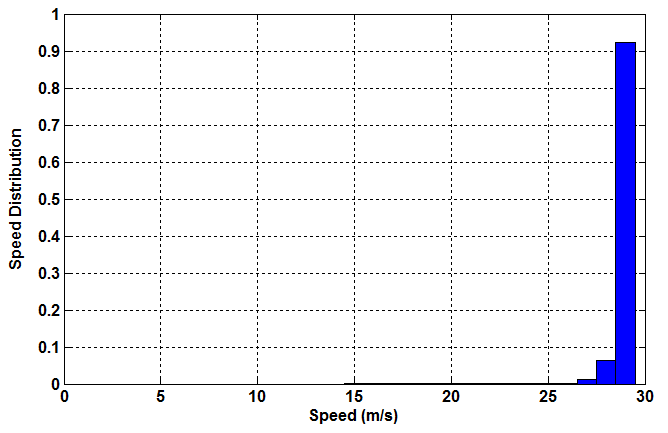}
  \includegraphics[width=0.52\textwidth]{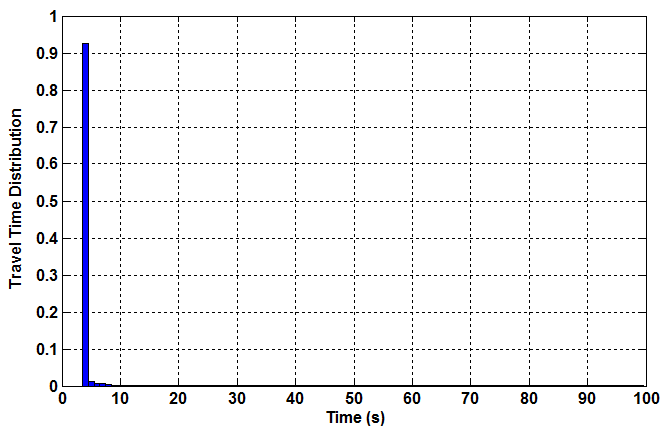}\\
    \caption{Car-speed probability distribution histogram, in left. Average travel time distribution histogram, in right. Model with downstream supply.
     The parameters of the road section are those of road section 1 in Table 1, and the arrival rate is $\lambda = 0.8$ veh/s.}
 \end{figure}
Fig. 10 displays the histograms for the probability distribution of the average car-speed and the average travel time through the road section, for our
model with downstream supply. The arrival rate is fixed to $\lambda= 2$~veh/s.

\begin{figure}[htbp]\label{fig-compar3}
  \includegraphics[width=0.56\textwidth]{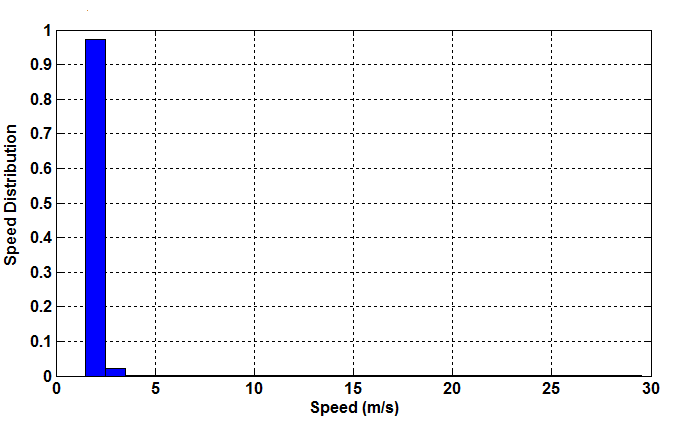}
  \includegraphics[width=0.5\textwidth]{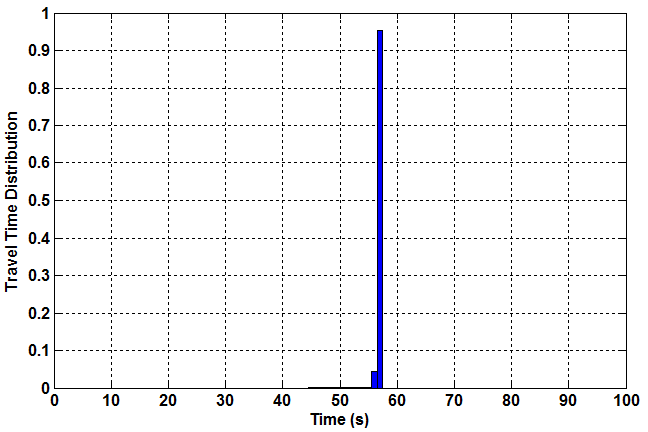}\\
    \caption{Car-speed probability distribution histogram, in left. Average travel time distribution histogram, in right. Model with downstream supply.
     The parameters of the road section are those of road section 1 in Table 1, and the arrival rate is $\lambda = 2$ veh/s.}
 \end{figure}

Fig. 10 shows that when the arrival rate $\lambda$ is large, the average car-speed is very low, and the average travel time is very
large (more than $3.5$ s).
Note that the average travel time in the road section is almost fifteen times larger than the free speed (about $60$ seconds).

\section{Conclusion and future work}

This paper presents a queuing model for road traffic that preserves the finite capacity property of the real system. Based on the $M/g/c/c$ state
dependent queuing model of Jain and Smith, we have proposed a stochastic queuing model for the road traffic which captures the stationary density-flow relationships in both uncongested and congestion conditions.

Experimental investigations of the proposed model are presented. Performance measures have been validated by comparison with $M/g/c/c$ state dependent
queuing model of Jain and Smith.
Car-speed and average travel time probability distributions are derived for two case of arrival rate.
The curves of those distributions shows that the proposed model correctly captures the interaction between upstream traffic demand and downstream
traffic supply.
Future work shall include the extension of the model to more than two sections in tandem to tree-topologies (complex series, merge, and split networks),
and consider the case where traffic demand, traffic supply and fundamental diagrams are stochastic.

 \section*{Acknowledgements}
The authors would like to thank the reviewers for their comments and suggestions that helped to improve this paper.

\bibliographystyle{plain}

\end{document}